\documentclass{ws-ijmpa}

\begin{document}

\markboth{A. Del Popolo, M. Gamnera, E.N. Ercan}
{Dynamical evolution of clusters of galaxies: the effect of high-velocity substructure clumps}

\catchline{}{}{}

\title{DYNAMICAL EVOLUTION OF CLUSTERS OF GALAXIES: THE EFFECT OF HIGH-VELOCITY SUBSTRUCTURE CLUMPS
\\
%USING \TeX\ OR \LaTeX\
}

\author{\footnotesize A. DEL POPOLO
}
\address{Dipartimento di Matematica, Universit\`{a} Statale di Bergamo,
  Piazza Rosate 2 - I 24129 Bergamo, ITALY} 
\address{Feza G\"ursey Institute, P.O. Box 6 \c Cengelk\"oy, Istanbul,
     Turkey}
\address{Bo$\breve{g}azi$\c{c}i University, Physics Department,
     80815 Bebek, Istanbul, Turkey}

\author{M. GAMBERA}
\address{Dipartimento di Matematica, Universit\`{a} Statale di Bergamo,
  Piazza Rosate 2 - I 24129 Bergamo, ITALY}

\author{E. NIHAL ERCAN}
\address{Bo$\breve{g}azi$\c{c}i University, Physics Department,
     80815 Bebek, Istanbul, Turkey
}

\maketitle

\pub{Received (Day Month Year)}{Revised (Day Month Year)}

\begin{abstract}
In the Cold Dark Matter (hereafter CDM) scenario even isolated
density peaks contain a high fraction of small scale clumps having
velocities larger than the average escape velocity from the
structure. These clumps populate protoclusters, especially in the
peripheral regions, $r \ge R_{\rm f}$ (where $ R_{\rm f}$ is the
filtering scale). During the cluster collapse and the subsequent
secondary infall, collapsing or infalling clumps (having $v<v_{\rm
esc}$) interact with the quoted {\it unbound} clumps (or {\it high
speed clumps}, as we also call them) having $v > v_{\rm esc}$. We
study the interaction between these two kind of clumps by means of
the impulse approximation$^1$ and we find that
the collapse of {\it bound} clumps is accelerated with respect to
the homogeneous case (Gunn \& Gott's model, Ref. 2). 
The acceleration of the collapse increases with
decreasing height of the peak, $\nu$. We finally compare the
acceleration produced by this effect to the slowing down effect
produced by the gravitational interaction of the quadrupole moment
of the system with the tidal field of the matter of the
neighboring proto-clusters studied in Del Popolo \& Gambera$^3$. We find
%%that nothwithstanding the two effects are totally different but
%%their
that
%%, only in the outskirts of the cluster,
the magnitude of the slowing down effect is larger than the
acceleration produced by the effect studied in this paper, only in
the outskirts of the cluster.
%%are comparable.
We want to stress that the one which we study in this paper is
also present in an isolated protocluster, being produced by the
interaction of the collapsing clumps with the {\it unbound}
substructure internal to the collapsing clumps itself
%and present in every density peak of the CDM
%scenario,
while that studied in Ref. 3 is produced
by substructure external to the density peak.
\end{abstract}

\section{Introduction}	%) A SECTION HEADING

According to the most promising cosmological scenarios,
structure formation is traced back to the evolution of primordial
density fluctuations. These fluctuations, originated from quantum 
fluctuations$^4$$^,$$^5$$ ^,$$^6$$^,$$^7$ in an inflationary
phase of the early Universe, grew up through gravitational
instability to a maximum radius $r_{\rm m}$. At the time $t_{\rm m}$
corresponding to maximum expansion, perturbations broke away from
the general expansion and at $\overline{\delta} \sim 1$ them began to
collapse. Hence the collapse of perturbations onto local density
maxima of the primordial density field has a key role in structure formation
and several studies deal with this problem. The problem of collapse has
been investigated from two points of view, namely:
\begin{description}
\item i) that of the statistical
distribution of the objects formed$^7$$^,$$^8$
\item ii) that of the structure of these objects and its dependence on the statistical
properties of the primordial density field$^2$$^,$$^9$$ ^,$$^{10}$$^,$$^{11}$  $^{12}$$^,$$^{13}$$ ^,$$^{14}$$^,$$^{15}$ $^{16}$$^,$$^{17}$$ ^,$$^{18}$$^,$$^{19}$ $^{20}$$^,$$^{21}$$ ^,$$^{22}$$^,$$^{23}$ $^{24}$.
\end{description}
In the spherical accretion model introduced by Hoyle and Narlikar$^{25}$
and applied to clusters of galaxies by Gunn \& Gott$^{2}$, the matter
around the core of the perturbation is a homogeneous fluid with
zero pressure. If the density inside the perturbation is greater
than the critical density, it is bound and shall expand to a
maximum radius $ r_{\rm m}$:
\begin{equation}
r_{\rm m} = \frac{r_i}{\overline {\delta}}
\end{equation}
where $r_{\rm i}$ is the initial radius and $ \overline{\delta}$ is
the overdensity inside the radius $ r$. Such matter shall collapse in a time
\begin{equation}
T_{\rm c0}/2 =\frac{ \pi}{ H_{\rm i}}
\frac{(1 +\overline{\delta})}
{\overline{\delta}^{3/2}}
\end{equation}
where $ H_{\rm i} $ is the Hubble parameter at the initial
time $ t_{\rm i} $. \\
This model was introduced in order to overcome the problem of the
excessively steep density profiles,
$\rho\propto r^{-4}$, obtained
in numerical experiments of simple gravitational collapse.
Some authors were able to produce shallower profiles$^2$$^,$$^{26}$$^,$$^{27}$,
$\rho\propto r^{-2}$, through the ${\it secondary}$ ${\it infall}$ process.
Several years later, observational
evidences for secondary
infall in the outskirts of clusters of galaxies
has been reported in Ref. 28, 29. \\
Even if
this model and in general
the original form of SIM (secondary infall model)
is able to explain better than previous models the
structure of clusters of galaxies it
has some drawbacks, (e.g.
it does not predict the right structure of density profiles), that
can be overcome with some improvements in the original model$^{30}$.
%
%This
%kind of problems can
%Slighth modifications of the model Improved
%version of SIM have a more larger predictive power ...(Del Popolo et al. 1999c)
Two noteworthy limits by the Gunn \& Gott$^{2}$ model are the following:
\begin{description}
\item {\it a}) It neglects tidal interaction of the perturbation with the neighboring
perturbation;
\item {\it b}) It neglects substructure existing in the outskirts of
proto-galaxies or proto-clusters
and in the background.
\end{description}
The effect of the tidal field was studied in a recent paper$^{3}$.
In that paper, we showed that the gravitational interaction of the
quadrupole momentum of a proto-structure with the tidal field of
the neighboring proto-structures introduces another potential
energy term in the equation describing the collapse which acts in
the sense to delay the collapse. This effect has revealed as one
of noteworthy importance in  solving several of the drawbacks of
the CDM model$^{3}$$^,$$^{31}$$^,$$^{32}$$^,$$^{33}$, namely:
\begin{description}
\item 1) lack of a mechanism originating the bias$^{3}$;
%Colafrancesco, Antonuccio \&
%Del Popolo 1995);
\item 2) discrepancies between:
\begin{itemize}
\item a1) X-ray temperature function of clusters and observations$^{31}$.
\item a2) two-point correlation
function of clusters
and observations$^{31}$;
\item a3) mass function and
velocity dispersion function of clusters and observations$^{33}$;
\item a4) shape of clusters and observations$^{34}$;
\item a5) angular two-point correlation function of galaxies and observations$^{32}$.
\end{itemize}
\end{description}
For what concerns point " $b)$", we may say that during the last
two decades, astronomers have discovered that at least a third of
galaxy clusters are not dynamically relaxed systems but contain
substructure on scales of the same order as the cluster itself$^{34}$$^,$$^{35}$$^,$$^{36}$. 
This implies that clusters are currently forming or have
formed recently enough that they have not had time to undergo
significant degree of violent relaxation and phase mixing. When
one studies the structure of the velocity field around a peak one
finds that at radii $ r$ larger than the filtering scale $ R_{\rm
f}$, ($r \ge R_{\rm f}$), a sensitive fraction of the matter of
the peak is gravitationally unbound to it$^{37}$. This is not surprising since a non zero
fraction of unbound matter should be expected even in the case of
a uniform system with a Maxwellian velocity distribution (see
section 2.2 for a discussion).

% Tolto
%%Moreover, high velocity stars are observed in the Galaxy
%%(Alexander 1982; Carney et al. 1988; Cudworth 1990, Leonard \&
%%Tremaine 1990), and also in the Local Group of galaxies (Valtonen
%%\& Wiren 1994).
%
The encounters between collapsing clumps and the
unbound high velocity clumps influence the collapse of the
proto-cluster. We want to study this point and to compare its
effect on cluster collapse with that produced by the effects of
tidal interaction
of the protocluster with the neighboring ones. \\
%and that of the substructure on the collapse of the perturbation
%are similar, they delays the collapse.
In the following we show how the unbound substructures influences
the collapse of the perturbation. The plan of this work is the following:
in Section ~2 we deal with the
unbound substructure effects. In Section ~3 we give our results
and finally Section ~ 4 is devoted to the conclusions.

\section{Unbound substructure effects}

Our cosmological setting is described by a hierarchical clustering
scenario, a CDM model (see Ref. 38) in which structure
formation goes {\it bottom-up}. Local density extrema are assumed
to be the sites of structure formation$^{7}$$^,$$^{8}$. The
perturbations grows linearly until the linearly evolved dispersion
$ \sigma(z,M)$ of the density contrast is $<1$, while it enters in
the non-linear phase when $ \sigma(z,M)=1$. Denoting this epoch by
$z_{\rm nl}$ one has:
\begin{equation}
\sigma_{0} (M) =1 +z_{\rm nl}
\end{equation}
(Liddle \& Lyth$^{38}$, equation 8.2), where as always the subscript
0 denotes the present value of the linearly evolved quantity and
%$ z_{\rm nl}$ is the redshift in the non-linear phase
%and
$\sigma_0(M)$ is related to the power spectrum $ P(k)$
of density fluctuations by:
\begin{equation}
\sigma_{0} (M) = \frac{ 1}{ 2 \pi^{2}} \int P(k) k^{2} W( k R_{\rm f}) d k
\end{equation}
being $ R_{\rm f} $ the filtering scale and
$ W( k R_{\rm f})$ the window function:
\begin{equation}
W( k R_{\rm f}) = \frac{ 3[ sin( k R_{\rm f}) -
k R_{f} cos( k R_{\rm f})]}{(k R_{\rm f})^{3}}
\end{equation}
A subgalactic mass scale collapses at a redshift:
\begin{equation}
1+z_{\rm nl} =\frac{30}{b}
\end{equation}
(Silk \& Stebbins$^{39}$), where $ b $ is the biasing parameter. In
particular a perturbation of $ 10^{15} M_{\odot} $ collapses at a
redshift $ z \simeq   0.02$ while perturbations in the range $
10^{6} M_{\odot}  \div  10^{9} M_{\odot}$ collapse almost at the
same $ z \simeq 18 $ because the mass variance in this region
varies only of a factor 3 (Ref. 40). A part of this last
perturbations have a cross section too little for gravitational
merging to be destroyed$^{40}$. Therefore a mass perturbation
$>10^8 M_{\odot}$ (see Ref. 38)
%
%%$ 10^{6} M_{\odot} $ \div $ 10^{9} M_{\odot}$
%
can survive until the
cluster enters non-linear phase at $ z\simeq 0.02$.\\
In order to study the dynamical evolution of the proto-cluster, we
use the shell method$^{3}$$^,$$^{41}$$^,$$^{42}$$^,$$^{43}$. If we
consider a shell of matter around the centre of the cluster this
is made of galaxies, gas and substructure. During the collapse,
the shell encounters clouds of substructure some of which have
high speed. Supposing that the systems subject to the encounter
(the shell of matter and the high speed clumps) have,
respectively, median radii $r_1$ and $r_2$, velocity dispersions
of order $\sigma_1$, $\sigma_2$, that at the instant of closest
approach their centres are separated by $b'$ and have relative
speed V the condition$^{44}$:
\begin{equation}
V>>\sigma_{\rm i} \frac{max(r_1,r_2,b')}{r_{\rm i}} \hspace{0.5cm} i=1,2
\label{eq:impuls}
\end{equation}
ensures that in the course of the encounter the objects
constituting the encountering systems barely move from
their initial locations.
%
%Clouds of substructure having
%velocity $ v >> \sigma $(where $ \sigma $ is the velocity dispersion
%of the galaxies and the substructure in the shell?)
%interact with the material contained in a collapsing shell and
The result of the encounter is to generate perturbations in the
velocity and energy of the objects constituting the shell (here we
are only interested in the effect of the encounter on the shell).
In fact the effects of high speed encounters on the internal
structure of "stellar" systems decrease as the encounter speed
increases and consequently high speed encounters generate only
small perturbations
of otherwise steady-state systems.\\
The conditions
%$ v >> \sigma $
expressed in equation (\ref{eq:impuls}) is known as impulse approximation (hereafter IA)$^{1}$ and
yields accurate results often even when the condition is not
strictly satisfied.
As stressed by Gnedin \& Ostriker$^{42}$, the IA works
well in the outer region of clusters, wherein galaxies are assumed to move
little during the interaction, while the IA is not appropriate
in the cluster core.

\subsection{Applicability of the IA}

Before going on, we want to analyze the applicability of the IA as
expressed by
%%%verify that the condition of
equation~(\ref{eq:impuls})
%%%is satisfied.
to our problem.

A first rough analysis can be performed remembering that the escape velocity in the cluster
is fixed by the virial theorem
%%Tolto
according to equation (see for example Ref. 44, (equation (8.3))):
\begin{equation}
<v^2_{\rm esc}>=4 <v^2>
\end{equation}
where $v_{\rm esc}$ is the escape velocity and $<v^2>$ is the 3-D dispersion
velocity,
and
\begin{equation}
<v^2>=3 \sigma_1^2
\end{equation}
being $\sigma_1$ the velocity dispersion of the objects
constituting the shell, and then
\begin{equation}
v_{\rm esc}=\sqrt{12} \sigma_1
\end{equation}
The relative velocity, V, is the sum of the infall velocity of the
shell and the velocity of clumps moving outwards:
\begin{equation}
V \simeq v_{\rm clumps}+v_{\rm infalling \hspace{0.5 mm} shell}
\end{equation}
Imposing that the clumps move at speed larger than $v_{\rm esc}$,
we get:
\begin{equation}
v_{\rm clumps}>v_{\rm esc}=\sqrt{12} \sigma_1 = 2 \sqrt{3}
\sigma_1 \label{eq:vee}
\end{equation}
The infalling shells are
marginally bound, hence they travel at a velocity close to the local escape
velocity
\begin{equation}
v_{\rm infalling \hspace{0.5 mm} shell} \leq v_{\rm esc}=
\sqrt{12} \sigma_1 \label{eq:vee1}
\end{equation}
This last statement can be easily proven using for example an
Hernquist model for the cluster. In general, conservation of
energy yields:
\begin{equation}
v_{\rm infalling \hspace{0.5 mm} shell}^2 = v_{\rm esc}^2 + 2
\Phi(r_{ a}),
\end{equation}
where $\Phi$ is the (negative) potential and $r_a$ is the maximum
(apocentric) radius of the infalling particle.

For example, for a Hernquist cluster, one has
\begin{equation}
\Phi(r) = - GM/(r+a)
\end{equation}
and
\begin{equation}
v_{\rm infalling \hspace{0.5 mm} shell}^2 (r)  = 2 [\Phi(r_a) -
\Phi(r)] = v_{\rm esc}^2(r) [1 - (r + a) / (r_a + a)] = v_{\rm
esc}^2(r) (r_a - r) / (r_a + a)
\end{equation}

For a Hernquist model, $a$ is a few tenths of the virial radius.
The infalling particles are between the virial and turnaround
radii, i.e. between $r_{\rm vir}$ and roughly 2 $r_{\rm vir}$. So
$ r_a>> a$ and
\begin{equation}
v_{\rm infalling \hspace{0.5 mm} shell}^2 (r)= v_{\rm esc}^2 (r)
(1 - r/r_a)
\end{equation}
which means that $v_{\rm infalling \hspace{0.5 mm} shell}$ is a
little smaller than $v_{\rm esc}$.

Using the previous results (equation~(\ref{eq:vee})and
(\ref{eq:vee1}) ), we have:
\begin{equation}
V \simeq v_{\rm clumps}+v_{\rm infalling \hspace{0.5 mm} shell}=
2 \sqrt{3} \sigma_1+\sqrt{12}\sigma_1 \simeq 7 \sigma_1
\end{equation}
If $i=1$,
in Eq. (\ref{eq:impuls}),
we are dealing with the shell
%%: $\sigma_1 \simeq \sigma$
then:  \\
1) $V=7 \sigma_1 > \sigma_1$ \\
If $i=2$, we are dealing with the clump: $\sigma_2 \simeq 1 {\rm
km/s} << \sigma_1$
(see Ref. 39) and then: \\
2) $V=7 \sigma_1>>\sigma_2$.  \\
Both the relations 1) and 2) satisfy the  condition:
\begin{equation}
V \geq 7 \sigma_1
\end{equation}
value that according to Ref. 44 (see page 440)
is considered good to use the IA.  \\
%%Although  the impulse approximation in the form of Eq. (\ref{eq:impuls})
%%is satisfied,
%%
We want to remark that the previous analysis does not take account of the reduction of
escape velocity with radius: given the potential of the sistem, $\Phi(r)$, the escape
velocity is $v_{\rm esc}=\sqrt{2 |\Phi(r)|}$. This means that going outwards the IA is more strictly satisfied.
%%In the case of peaks having $\nu=1.2-2$ the escape velocity from the cluster,
%%$v_{\rm esc}$, is $> (1.5  \div  2) \sigma_1$ in
%%the range $(0  \div  7) h^{-1}$ Mpc (Antonuccio-Delogu \& Colafrancesco 1995),
%%where $\sigma_1$ is the cluster velocity dispersion.
%%We will consider only high speed clumps characterized by
%%$ v \geq 3 v_{\rm esc} > (4.5  \div  6) \sigma_1$ in order the impulse
%%approximation to be satisfied (see Binney and Tremaine 1987).

%%have eliminated the unknown
%%coming from the applicability della 7?????

\subsection{Energy change in the shell}

The energy change in the shell, due to an encounter
with high velocity clumps, can be computed as follows. \\
%Suppose that the
%{\it A}
Suppose to have a perturber (a high speed clump of substructure,
member of the cluster) characterized by a speed larger than the
escape speed from the cluster. This high velocity clump, that we
stress is inside and part of the cluster, moves away from it. We
use cylindrical coordinates (R,z) such that the z-axis coincides
with the perturber's trajectory given by:
\begin{equation}
R=0, \hspace{0.5cm} z = A(t) = Vt
\end{equation}
and that $ R = 0, z = 0$ is the centre of the perturbed system
(shell of matter). Note that we are interested only in clumps not
contained in the core and moving from the centre in the outward
direction, so that they do not go through the cluster centre. The
choice of not taking into account the effect of clumps in the core
area is due to Gnedin \& Ostriker$^{42}$ observation that the IA
works well in the outer region of clusters, wherein galaxies are
assumed to move little during the interaction. The velocity
increment $ \Delta {\bf v}$ of a test particle at the position
(R,z) is given by:
\begin{equation}
\Delta v_{\rm R} = \int_{0}^{\infty} \frac{d \Phi}{dr} \frac{R}{r}dt =
- \frac{R}{V} \int_0^{\infty} \frac{d \Phi}{dr} \frac{dA}{r}
\end{equation}
where
\begin{equation}
r = [(A - z)^2 + R^2]^{1/2}
\end{equation}
and $ \Phi$ is the potential of the perturber. Assuming that the
perturber is an isothermal sphere it can be easily shown that:
\begin{equation}
\Delta v_{\rm R}=-\pi \sigma_2^2/V
\label{eq:iso}
\end{equation}
where $\sigma_2$ is the perturber internal velocity dispersion. \\
The energy, gained by the shell of matter during the encounter
with a high velocity clump, is given by:
\begin{equation}
\Delta E = \int_{0}^{\infty} [ \Delta v(R)]^2 \Sigma(R) R dR
\label{eq:ene}
\end{equation}
(see Ref. 44) where $ \Sigma(R)$ is the surface density of
the perturbed system.

We assume that
%the perturber is an isothermal sphere
%and
the perturbed system (shell) is characterized by an average density
given by:
\begin{equation}
\overline{\rho}(r_{\rm i},t)=\frac{3 M}{4 \pi a^3(r_{\rm i},t) r^3_{\rm i}}
\label{eq:mass}
\end{equation}
where, according to Ref. 2 notation, $a(r_{\rm i},t)$ is the expansion
factor of the shell at time $t$, $r_{\rm i}$ its initial radius
and $M$ its mass. We are studying the collapse of a shell of
matter then
%(note that for a collapsing shell
$t_{\rm i}$ and
$r_{\rm i}$ coincide with the turn-around epoch and radius of the shell,
respectively.
Assuming that no shell crossing occurs,
the total mass inside each shell stays constant and then:
\begin{equation}
\overline{\rho}(r_{\rm i},t)=\frac{\overline{\rho}(r_{\rm i},t_{\rm i})}{a^3(r_{\rm i},t)}
\label{eq:mas}
\end{equation}
The surface density is given by:
\begin{equation}
\Sigma=\frac{\Sigma_{\rm i}}{a(r_{\rm i},t)^2}
\label{eq:surf}
\end{equation}
%%
%%%a De Vacouleurs profile having surface density:
%%\begin{equation}
%%\Sigma = \Sigma_0 \cdot e^{- 7.67(\frac{R}{R_e})^{0.25}}
%%\end{equation}
%%where $ R_e$ is the effective radius of the shell.
At this point we are ready to calculate
the energy change in the shell
%is then given,
by using
equations~(\ref{eq:iso}), (\ref{eq:ene}), (\ref{eq:surf}) and
remembering that $r=r_{\rm i} a(r_{\rm i},t)$:
\begin{equation}
\Delta E= \pi^2 \frac{\sigma_2^4}{V^2} \Sigma_{\rm i} r_{\rm i}^2
(\log a_{\rm max}-\log a_{\rm min})
\end{equation}
where $a_{\rm max}$ is the maximum value of the expansion
parameter, corresponding to the turn-around epoch, and $a_{\rm
min}$ the minimum value of the expansion parameter corresponding
to the time of full collapse of the perturbation.
%
%%
%%
%%
%%\begin{equation}
%%\Delta E= \pi^2 \frac{\sigma_2^4}{V^2} \Sigma_{\rm i} r_{\rm i}^2
%%\log \frac{R^{\ast}_{\rm min}}{R_{\rm c}}
%%\end{equation}
%%In the following, we assume that the cut off radius $R_{\rm c}=0.1
%%R_{\rm min}$.
%
%%%
%%%Spiegare l'introduzione del cut off..
%%%
%%
%%%where $a_{\rm max}$
%%%is the maximum value of the expansion parameter,
%%%corresponding to the turn-around epoch,
%%%and $a_{\rm min}$
%%%the minimum value of the expansion parameter
%%%corresponding to the time of full collapse of the perturbation.
%%
%%
%%\begin{equation}
%%\delta E = \frac{\Sigma_0}{(7.67)^8} \left(\frac{4 \pi \theta_{1}^{2} R_e}{V} \right)^{2} \int_{0}^{\infty} n^{7} e^{-n} dn
%%\end{equation}
%%where $\theta_{1}$ is the velocity dispersion of the perturber and
%%\begin{equation}
%%\Sigma_0 = 2n_{0}R_{core} = \frac{ 9 \theta_{2}^{2} }{2 \pi G R_e}
%%\end{equation}
%%where $ \theta_{2}$ is the velocity dispersion of the perturbered
%%system and $ R_e$ is its core radius.\\
During infall a shell of matter experiences several encounters
with all the unbound clumps located between its initial position
and the final one. Each one of the encounters produces an increase
of the energy of the shell. In order to calculate the total
energy, $\Delta E_{\rm T}$, acquired during infall we must
calculate the number of unbound clumps "internal" to its orbit.
The number of perturbers can be calculated observing that:
\begin{description}
\item aa) they are peaks of the density field having dimension
much less than that of the cluster
but large enough to survive till the cluster enters the non-linear phase;
%in order to minimize the
%tidal field effects than they can produce.
\item bb) they must satisfy the IA condition.
\end{description}
Clumps in the SCDM scenario are not distributed randomly but are
clustered and so
tend to be more bound to their nearby neighbors. This may make them more
subject to mergers in some cases soon after they form, but may also
prevent mergers by placing them in a high velocity dispersion environment.
For survival, we must require a large density contrast to have developed.
\begin{figure}
\psfig{file=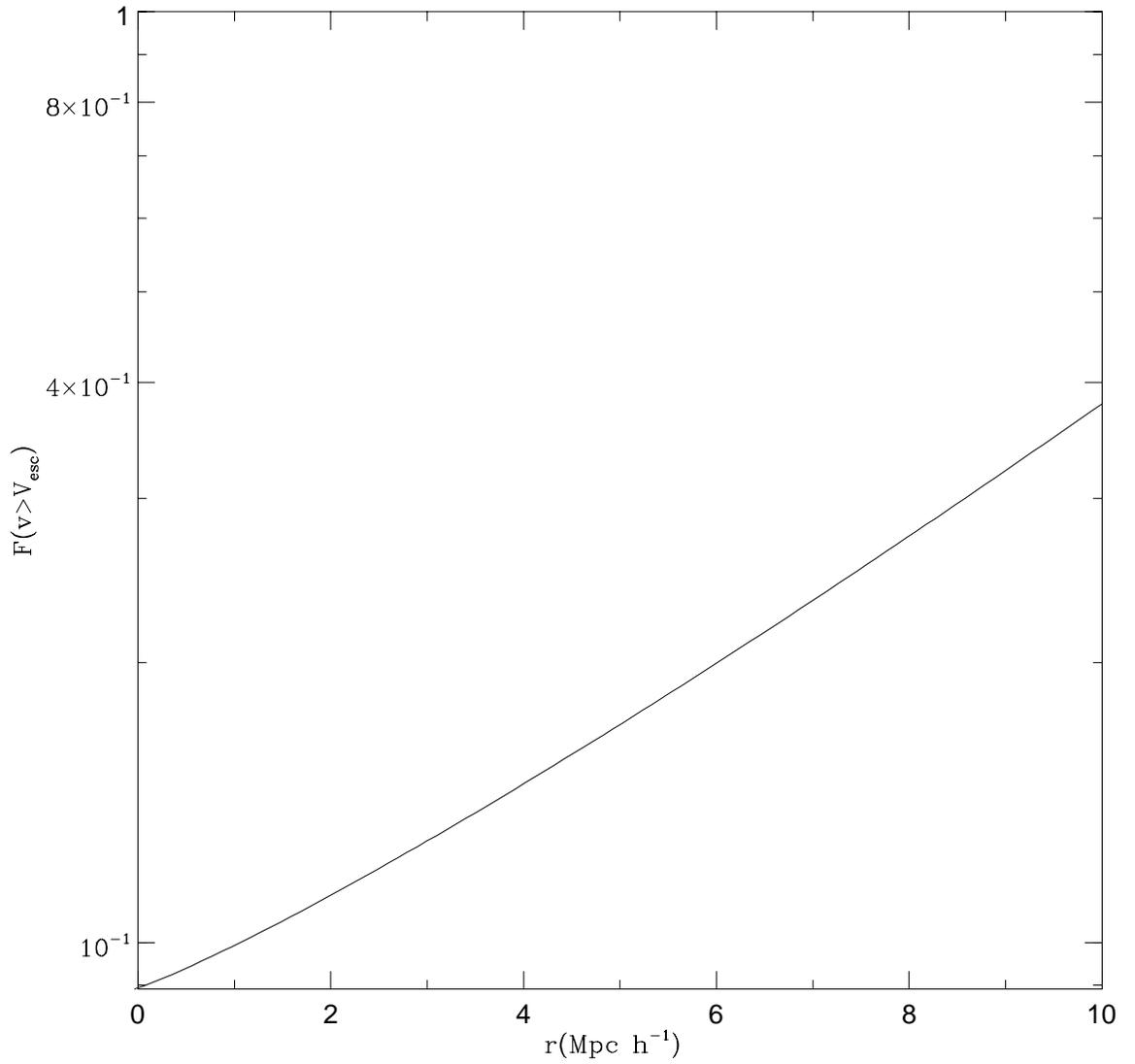,width=16cm}
\caption[]{Spatial variation of $F (v \geq v_{\rm esc})$
for the CDM spectrum, introduced in Section ~2,
smoothed on a scale
$R_{\rm f} =5 h^{-1}$ Mpc and for $\nu=1.2$.
}
\label{Fig. 1}
\end{figure}
If the smaller scale fluctuations have a density contrast of $\simeq 100$, they
should be able to survive tidal disruption or dynamical friction drag
when its environment undergoes non-linear collapse. This translates
into a minimum mass scale of $ \simeq 10^8 M_{\odot}$ (see Ref. 39).
In order to determine the number density and the characteristics of these
peaks we follow the same choice made by Antonuccio-Delogu \& Colafrancesco$^{44}$
and Del Popolo \& Gambera$^{43}$. We restrict ourselves to consider a subset
of peaks of the density field, having the central height $\nu$ larger
than a threshold $\nu_{\rm c}$. The threshold is chosen by
satisfying the condition its radius $r_{\rm pk}$ is smaller than the
average peak separation, $d_{\rm av}=n_{\rm a}^{-1/3} ( \nu \geq \nu_{\rm c})$:
%The condition a) can be written in the form:
\begin{equation}
r_{\rm pk}( \nu \geq \nu_{\rm c}) \le 0.1 \cdot n_{\rm a}^{-1/3}
( \nu \geq \nu_{\rm c})
\label{eq:asterisco}
\end{equation}
where the radius of a peak $ r_{\rm pk}$ is given by:
\begin{equation}
r_{\rm pk} = \sqrt{2} R_{\star} \left[ \frac{1}{1 + \nu \sigma_{0}} \cdot
\frac{1}{\gamma^3 + (0.9\nu)^{1.5}} \right]^{\frac{1}{3}}
\end{equation}
(see Ref. 45), where:
\begin{equation}
\gamma=\sigma^2_1/\sigma_0 \sigma_2
\end{equation}
and here $\sigma_{\rm i}$ is the i-th momentum of the variance of the
fluctuation field:
\begin{equation}
\sigma^2_{\rm i}=\frac{1}{2 \pi^2} \int P(k) k^{2(i+1)} dk
\end{equation}
where $P(k)$ is the filtered power spectrum. The power spectrum
that we adopt is $P(k)=Ak T^2(k)$ with the transfer function
$T(k)$ given in Ref. 7 (equation~(G3)):
%\newpage
\begin{equation}
T(k) = \frac{[\ln \left( 1+2.34 q\right)]}{2.34 q} \cdot [1+3.89q+
(16.1 q)^2+(5.46 q)^3+(6.71)^4]^{-1/4}
%
%T^2(k) &=& [\ln \left( 1+4.164k\right)]^2 \cdot (192.9+1340k+ \nonumber \\
%& + &  1.599\cdot 10^5k^2+1.78\cdot 10^5k^3+3.995\cdot
%10^6k^4)^{-1/2}
%
\label{eq:ma5}
\end{equation}
where $ A$ is the normalizing constant and $q=\frac{k
\theta^{1/2}}{\Omega_{\rm X} h^2 {\rm Mpc^{-1}}}$. Here
$\theta=\rho_{\rm er}/(1.68 \rho_{\rm \gamma})$ represents the
ratio of the energy density in relativistic particles to that in
photons ($\theta=1$ corresponds to photons and three flavors of
relativistic neutrinos).
%
%%Here $P(k)$ is the filtered power spectrum.
%%For a CDM model with spectrum given by:
%%\begin{eqnarray}
%%P(k)&=& k^{-1} [\ln \left( 1+4.164k\right)]^2 \cdot (192.9+1340k+ \nonumber \\
%%& + & 1.599\cdot 10^5k^2+1.78\cdot 10^5k^3+ \nonumber\\
%%& + & 3.995\cdot
%%10^6k^4)^{-1/2}
%%\label{eq:ma5}
%%\end{eqnarray}
%%(Ryden \& Gunn 1987; BBKS)
For the previous CDM transfer function, normalized as
$\sigma_8=0.63$ and filtering radius $R_{\rm f}=30 {\rm Kpc}$,
(corresponding to a Gaussian filtered mass $M \simeq 10^8
M_{\odot}$) one has $\gamma =0.48$, $R_{\ast}=\sqrt{3
\frac{\sigma_1}{\sigma_2}}=33Kpc$ and then $ \nu_{c} = 1.6$.
\begin{figure}
\psfig{file=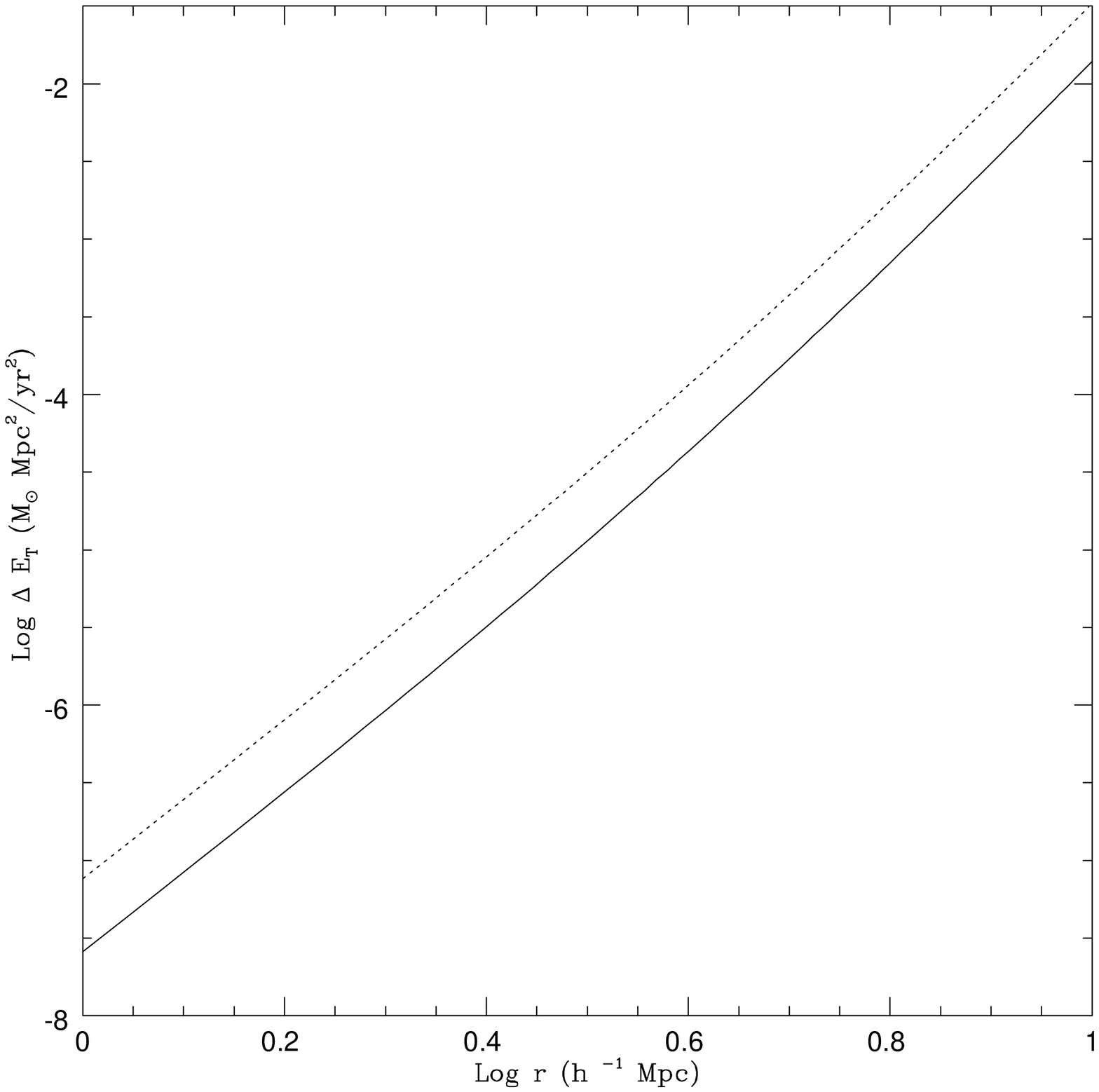,width=16cm}
\caption[]{Energy acquired during encounters of
shells of matter infalling from a distance $r$ from the centre of
the proto-cluster. The dashed line represents $\Delta E_{\rm T}$
for the CDM spectrum, introduced in Section ~2, smoothed on a
scale $R_{\rm f} =5 h^{-1}$ Mpc and for $\nu=1.2$. The solid line
is the same but now $\nu=2$} \label{Fig. 2}
\end{figure}
Remembering that the number
density of peaks, $n_{\rm a}$,
having $\nu \ge \nu_{\rm c}$, is given by:
\begin{equation}
n_{\rm a}=\int_{\nu_{\rm c}}^\infty N_{\rm pk} d \nu
\end{equation}
where
\begin{equation}
N_{\rm pk}=\frac{1}{(2 \pi)^2} \frac{1}{R_{\ast}^3}
G(\gamma, \gamma \nu) \exp(-\nu^2/2)
\end{equation}
being $G(\gamma, \gamma \nu)$ defined in Ref. 7 (equation~(4.4))
together with $R_{\ast}$ and $\gamma$, we obtain $n_{\rm a}
\simeq 135 {\rm Mpc}^{-3}$
%
%%\simeq 72 {\rm Mpc}^{-3}$.
%
%Eq. (\ref{eq:asterisco}) expresses the condition that
%the typical size of peaks is much less than their separation.
%
%As
%a consequence of this assumption,
%the contribution to the total energy
%acquired by a shell during encounters with unbound clumps is underestimated
%being these only a subset of the overall peak population. \\
%Using the expression for the upcrossing points given by BBKS
%(Eq.  ... ):
%\begin{equation}
%$n_a =
%\end{equation}
%$ n_a \sim 72 Mpc^{-3}$.
The average mass of
the peaks with $\nu \ge \nu_{\rm c}$ is given by:
\begin{equation}
m_{\rm a} = \frac{1}{n_{\rm a}( \nu \geq \nu_{\rm c})}
\int_{\nu_{\rm c}}^{\infty}
m_{\rm peak} N_{\rm pk}(\nu)d\nu \simeq 7 \times 10^8 M_{\odot}
\end{equation}
where
\begin{equation}
m_{\rm peak}=\frac{4 \pi}{3} \rho_{\rm b} R^3_{\ast}
\frac{2^{3/2}}{\gamma^3+(0.9/\nu)^{3/2}}
\end{equation}
(see Ref. 46). This mass satisfies both condition "
aa) " and
Silk \& Stebbins$^{39}$ prescription for surviving clumps.\\
The fraction of this subpeaks
%having a velocity $ v >> \sigma$ larger
%than the
%velocity dispersion in the shell,
satisfying the IA, and consequently satisfying condition " bb)",
can be calculated using the conditional probability distribution $
f_{\rm pk}({\bf v}|\nu) $ giving the peculiar velocity around a
peak of a gaussian density field having central height $\nu$:
\begin{equation}
f_{\rm pk}({\bf v}|\nu) =  \frac{\frac{\alpha}{4\pi^2}}{\sqrt{\pi^3
\lambda_{\rm i}}}
\cdot e^{- \sum_{i=1}^{3} \frac{[v_{\rm i} - \nu \langle \nu v_{\rm i} \rangle]^2}{\lambda_{\rm i}}}
\end{equation}
(see Ref. 37) ($<>$ denotes ensemble averages),
where
\begin{equation}
\alpha = \frac{\sqrt{ 2}}{\pi} erfc \left[\frac{\nu \langle \nu x \rangle}{\sqrt{\lambda x}} \right]
\end{equation}
$\langle \nu x \rangle=\gamma$, (Ref. 7)
\begin{equation}
\langle \nu v_{\rm i} \rangle=-\frac{H_0 \Omega_0^{0.6} r_{\rm i}}{r^3 \sigma_0} \int_0^r d x x^2 \xi(x)
\end{equation}
$\xi(r)$ is the autocorrelation function
and
\begin{equation}
x = - \frac{\bigtriangledown^2 \delta}{\sigma_2 |_0}
\end{equation}
is the central curvature of the peak. $\lambda_1$, $\lambda_2$, $\lambda_3$ are
three of the four eigenvalues of the covariance matrix $M$ described in
Antonuccio-Delogu \& Colafrancesco$^{37}$.
%..... values of the velocity correlation tensor
These three eigenvalues represent the components of
the peculiar velocity dispersions along directions parallel ($\lambda_1$)
and perpendicular ($\lambda_2, \lambda_3$) with respect to the radial
direction of a density peak.
%%the axes of the density peak.\\
%%
%%We study the effect of substructure on the spherical collapse model, then
%%we suppose that the velocity distribution around the peak is spherical that is:
%%\begin{equation}
%%f(v) =  A \cdot e^{- \frac{v^2}{2 \theta^2}}
%%\end{equation}
%%with this hypotesis the effect of non-radial motion are negleted. These
%%motions (Del Popolo \& Gambera 1998a,b,c,d) produce a delay of the
%%collapse increasing the substructure effects that we are going to calculate.
%%%
%%In the case of peaks having $\nu=1.2  \div  2$ the escape velocity from
%%the cluster, $v_{\rm esc}$, is $ > (1.5  \div  2) \sigma_1$ in
%%the range $(0  \div  7) h^{-1}$ Mpc (Antonuccio-Delogu \& Colafrancesco 1995),
%%where $\sigma_1$ is the cluster velocity dispersion.
%%
%%
%%%We will consider only high speed clumps characterized by
%%%$ v \geq v_{\rm esc}$
%%%in order the impulse
%%%approximation to be satisfied (see Binney and Tremaine 1987 and the discussion
%%%in the following).
%%
The fraction of peaks having $ v \geq v_{esc}$
%
%where
%\begin{equation}
%v_{esc} =  3 \sqrt{2 \Phi(r)} \simeq 6 \sigma
%\end{equation}
%is given by
can be obtained by means of the conditional
probability distribution $f_{\rm pk}(v|\nu)$:
\begin{equation}
F (v \geq v_{\rm esc}) = \frac{ \int_{|v| \ge v_{\rm esc}}^{\infty}
f_{\rm pk}(v|\nu)d^{3}  v}{\int_0^{\infty} f_{\rm pk}(v|\nu)d^{3} v}
\end{equation}

The fraction of peaks satisfying the condition $F (v \geq v_{\rm
esc})$ can be easily calculated to be $F (v \geq v_{\rm
esc})=0.00738$ for a Maxwellian velocity distribution, $f(v)
\propto \exp(-\frac{v^2}{2 \sigma^2_{\rm v}})$, characterized by a
uniform velocity dispersion $\sigma^2_{\rm v}$. In the case of a
top-hat perturbation of radius $R_{\rm f}=1.5 h^{-1} {\rm Mpc}$,
$F (v \geq v_{\rm esc})$ can be as large as $0.3$ for a typical
velocity dispersion, and even larger for larger values of $R_{\rm
f}$.
% Tolto
%%%%%%%%%%%%%%%%%%%%"Rivedere quello tolto"\\
%%We want to remark that in the case of the Galaxy, high
%%velocity stars are observed and used to infer the local galactic
%%escape speed (Alexander 1982; Carney et al. 1988; Cudworth 1990,
%%Leonard \& Tremaine 1990), and also in the Local Group of galaxies
%%some high speeds are observed (Valtonen \& Wiren 1994). In
%%particular Leonard \& Tremaine (1990) got the fraction $F$ of
%%stars in a given sample  with apparent space velocity $v^{'}$ in
%%excess of $v_{\rm esc}$. They obtained a value of $F (v \geq
%%v_{\rm esc})=0.26$, for $k=z=1$, and $F (v \geq v_{\rm
%%esc})=0.155$ for $k=1$, $z=1.5$, where $k$ is the power law index
%%of the distribution of high-velocity stars near the Sun and
%%$z=\frac{v_{\rm esc}-v^{'}_{\rm cut}}{\sigma_{\rm v }}$, where
%%$v^{'}_{\rm cut}$ is the apparent low-velocity cutoff in $v^{'}$
%%of the sample.
%

Using the conditional probability distribution $ f_{\rm pk}({\bf
v}|\nu) $, in the case of a
%%a larger value for the fraction of peaks.
%For example
%%
%%%while for
%%%$v>v_{\rm esc}$, in the case of a Maxwellian distribution,
%%%we get $F (v \geq v_{\rm esc})=0.00738$
%%
$\nu=1.2$ peak we get $F (v \geq v_{\rm esc}) \simeq 0.09-0.13$
within $R_{\rm f}=5 h^{-1} {\rm Mpc}$.

\footnote{It is important to note that the use of the peak formalism is not 
fundamental for what concerns the results of this paper. Similar results are 
obtained even using a top-hat perturbation for which, has previously remarked, 
$F (v \geq v_{\rm esc})$ can be as large as $0.3$.}
%
%%of the same order of magnitude as that derived by Tyson \& Fisher
%%(1995) from gravitational lensing observations.
In Fig. 1, we plot the spatial variation of $F (v \geq v_{\rm
esc})$ for the CDM spectrum, introduced in Section ~2, smoothed on
a scale $R_{\rm f} =5 h^{-1}$ Mpc and for $\nu=1.2$. In any case
in the following, we'll make a conservative calculation using the
lower estimates for $F (v \geq v_{\rm esc}) \simeq 0.09$.
%given
%the lower value for it
%Tolto
%In any case in the following, we'll make a conservative
%calculation using the lower estimates for $F (v \geq v_{\rm esc})$ given
%by the Maxwellian distribution.
%
%%
%%%The reason of this choice is due to a
%%%possible obiection to the applicability of Eq. (\ref{eq:impuls}).
%%

Finally, the total number of subpeaks verifying the above
conditions " aa) ", " bb) " is given by:
\begin{equation}
N_{\rm tot} = \frac{4 \pi}{3} r_{\rm i}^3 \cdot n_{\rm a} \cdot
F (v \geq v_{\rm esc})
\label{eq:num}
\end{equation}
while the total energy, $ \Delta E_{\rm T} $, injected in the shell
by
\begin{figure}
\psfig{file=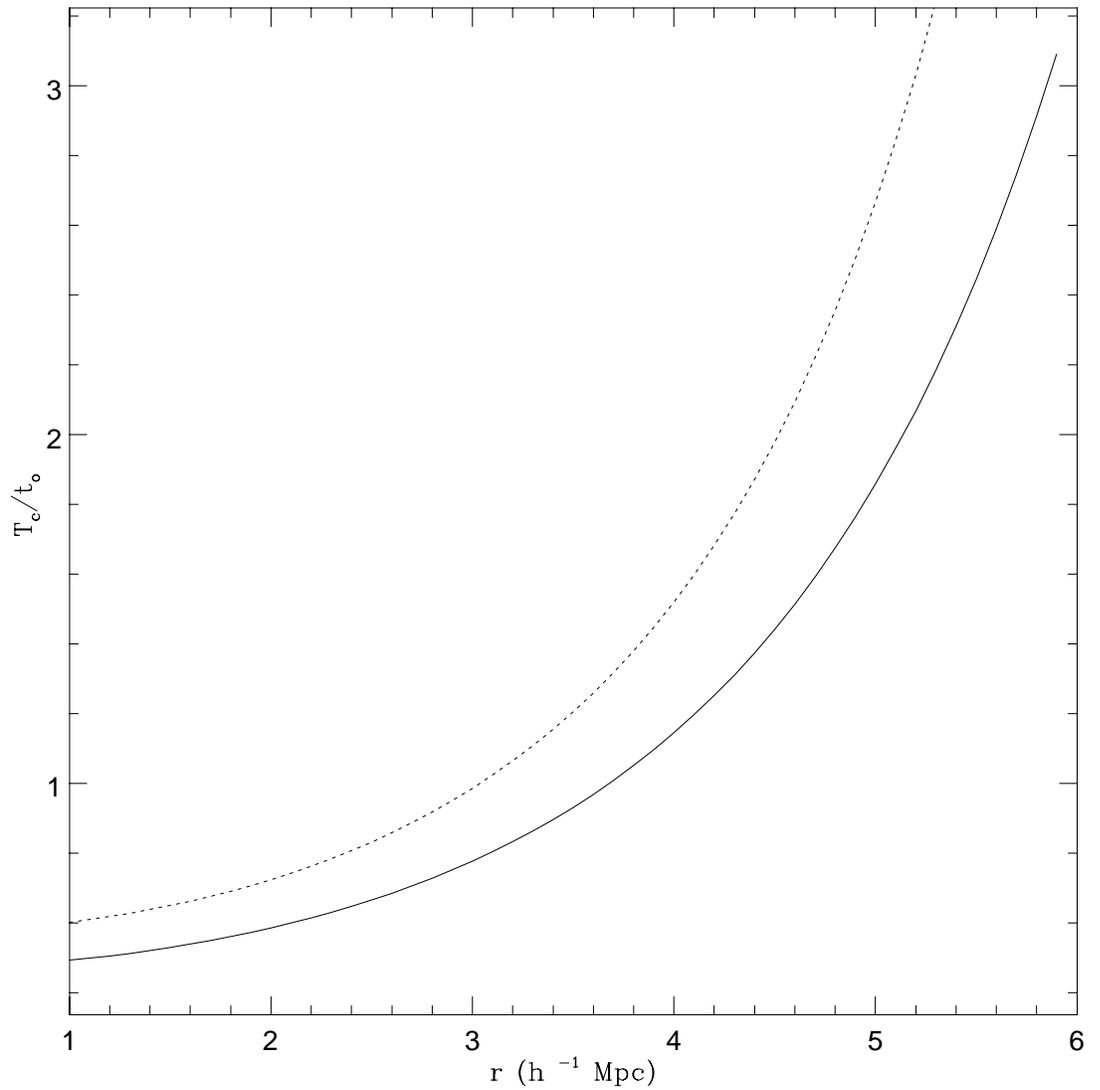,width=16cm}
\caption[]{The time of collapse of a shell of
matter around a peak having $\nu=1.2$ in a CDM model taking
account of the effect of the unbound high velocity clumps (solid
line) and for the Gunn \& Gott$^{2}$ model (dashed line). $t_{\rm o}$ is Hubble
time.} \label{Fig. 3}
\end{figure}
the impulsive encounters
with all high speed subpeaks is given by:
%the impulsive encounters, $ \Delta E_{\rm T} $,
%produced by the encounters
%with all unbound subpeaks is given by:
\begin{equation}
\Delta E_{\rm T} = N_{\rm tot} \cdot \Delta E
\end{equation}
The energy for unit mass of a collapsing shell is a modified version of that
given by Ref. 47 (equation(~19.9)) and can be written similarly
to Ref. 48 in the form:
%equation given by Peebles (1993):
\begin{equation}
E = \frac{1}{2} \left(\frac{dr}{dt}\right)^2 - \frac{GM}{r} + \Delta E_{\rm T}
\label{eq:ener}
\end{equation}
where $ M$ is the mass of the shell of matter.\\
%and $ r$ its radius. \\
Integrating equation~(\ref{eq:ener}), using
equations~(\ref{eq:mass}), (\ref{eq:mas}) and
remembering that $\rho_{\rm i} =\rho_{\rm ci} (1+\overline{\delta})$ where
$ \rho_{\rm ci} = \frac{3H_{\rm i}^2}{8\pi G}$, with
$\rho_{\rm ci}$ and $H_{\rm i}$ respectively
the critical mass density and the Hubble constant at the time $t_{\rm i}$,
the time of collapse is given by:
\begin{equation}
t_{\rm c}=2 \int_0^{a_{\max }}\frac{da}{\sqrt{H_{\rm i}^2
\left[ \frac{1+\overline{\delta} }a-
\frac{
1+\overline{\delta }}{a_{\max }}\right]
-\frac{4 G}{H_{\rm i}^2 (1+\overline{\delta }) r_{\rm i}^5 }
\Delta E_{\rm T}}}
\label{eq:tco}
\end{equation}
Integrating analytically equation~(\ref{eq:tco}), we find the
solution:
\begin{equation}
t_{\rm c}/2 \simeq \frac{ \pi}{ H_{\rm i}}
\frac{1}{\left(
\overline{\delta}+
\frac{4 G}{H_{\rm i}^4 r_{\rm i}^5 }
\Delta E_{\rm T}
\right)^{3/2}}
\end{equation}
having assumed $\overline{\delta }<<1$. In the case $\Delta E_{\rm
T}=0$ the solution of this equation is that given In Ref. 2:
\begin{equation}
t_{\rm c0}/2 \simeq \frac{ \pi}{ H_{\rm i}}
\frac{1}
{\overline{\delta}^{3/2}}
\end{equation}
%%
%%\begin{equation}
%%T =  \int_{0}^{R} \frac{dr}{\sqrt{2(E + \frac{GM}{r} -  N_{tot} \Delta E)}}
%%\end{equation}
%%using the virial theorem we have:
%%\begin{equation}
%%T =  \int_{0}^{R} \frac{dr}{\sqrt{ \frac{GM}{r} -  N_{tot} \Delta E}} \simeq
%%T_0 \left(1 + \frac{3  N_{tot} \Delta E R}{10 GM} \right)
%%\end{equation}
%%where
%%\begin{equation}
%%T_0  = \frac{2}{3} \cdot \frac{R^{1.5}}{\sqrt{ GM}}
%%\end{equation}

\section{Results and discussion}

The results found by means of the model introduced in the
previous section are shown in the Fig. 1-5.

As previously reported, Fig. 1 shows the spatial variation of $F
(v \geq v_{\rm esc})$ for the CDM spectrum, introduced in Section
~2, smoothed on a scale $R_{\rm f} =5 h^{-1}$ Mpc and for
$\nu=1.2$. The fraction of clumps with $v \ge v_{\rm esc}$ within
$R_{\rm f}= 5 h^{-1} {\rm Mpc}$ is $F (v \geq v_{\rm esc}) \simeq
0.09-0.13$, and increases with increasing radius. This behavior is
due to the fact that the potential of the structure becomes
shallower going towards its outskirts, and consequently $v_{\rm
esc}$ is substantially lower. In Fig.~2 we show the total energy
injected in the shell, because of the impulsive encounters of
collapsing shells with the high speed clumps, in function of the
initial radius of the shell. The dashed line represents $\Delta
E_{\rm T}$ for a CDM model with spectrum given by
equation~(\ref{eq:ma5}) for $\nu=1.2$. As shown, shells collapsing
from larger distances from the centre of the cluster acquire more
energy. This is due to two different reasons: firstly the energy,
$\Delta E$, acquired in an encounter with a single clump increases
with increasing turn-around radius of the shell of matter.
%
%%: if the turn-around radius is larger
%the shell starts its collapse from a larger distance from the centre
%of the cluster and consequently it shall encounter more clumps.
%
%(meglio raggio massimo, aggiustare l'equazione
%dell'energia). ............;
Secondly, $\Delta E_{\rm T}$, is proportional to the number of
unbound clumps that increases with radius (see
equation~(\ref{eq:num})) and then
%because
%the fraction of unbound clumps, $F(v \ge v_{\rm esc})$, increases with
%radius and because
a shell falling from a larger distance encounters
more unbound clumps.
\begin{figure}
\psfig{file=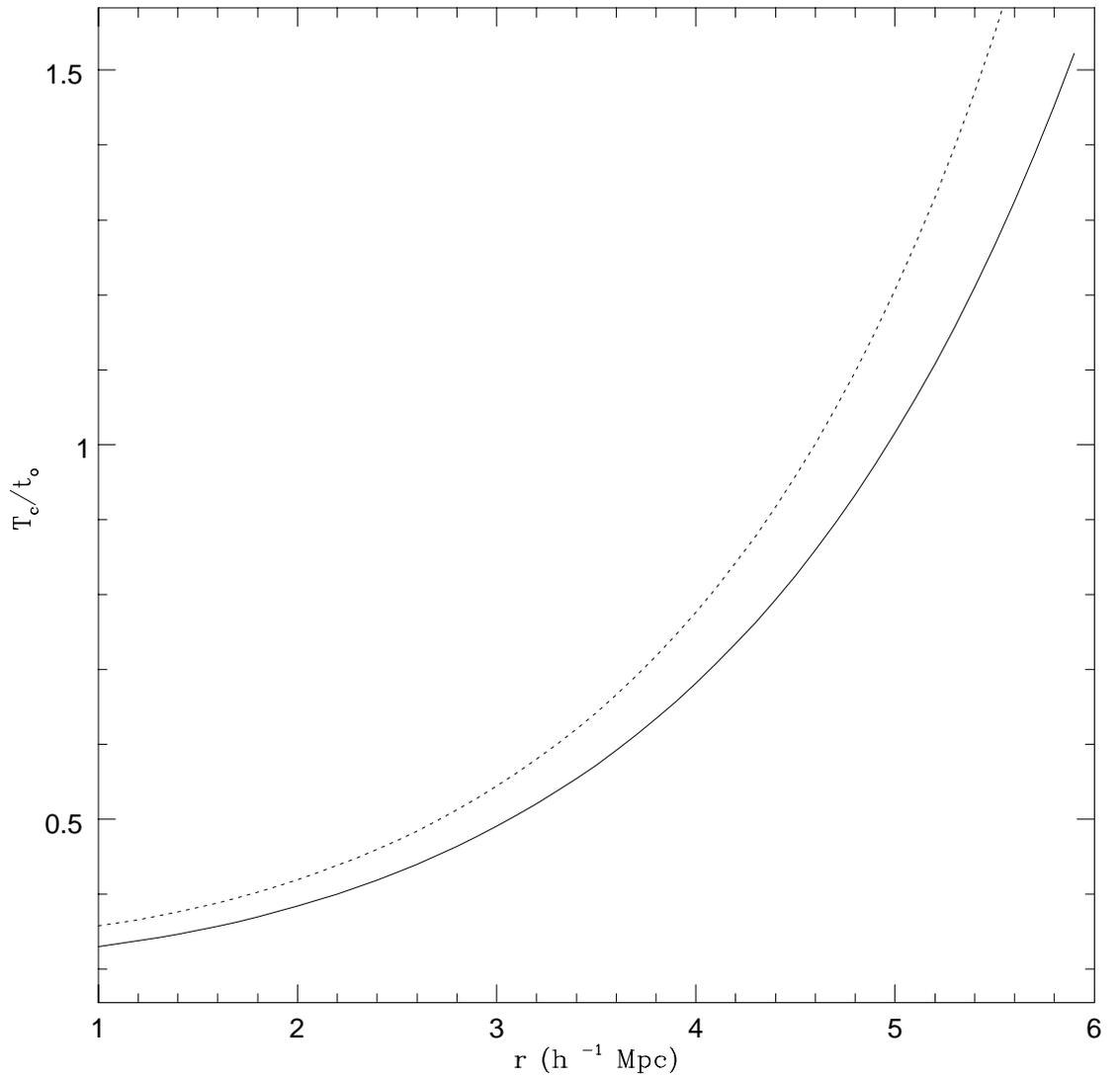,width=16cm}
\caption[]{Same as Fig.~3 but now $\nu=2$}
\label{Fig. 4}
\end{figure}
The solid line represents $\Delta E_{\rm T}$ for $\nu=2$. The value
of $\Delta E_{\rm T}$ is smaller with respect to the previous case. The
reason is due to the fact that the fraction of the unbound peaks
decreases for higher $\nu$ peaks because the potential of these protostructures
is deeper and hence the value of $v_{\rm esc}$ is systematically higher. \\
In Fig.~3 we show the collapse time versus the distance from the centre
of the cluster. The figure plots the collapse time from 1 $h^{-1}$${\rm Mpc}$ on,
because
of the limits of the IA, stressed by
Ref. 42, in the
central part of the cluster.
The solid line represents, $t_{\rm c}$, for $\nu=1.2$ and
taking account of the unbound clumps effect, while
the dashed line is the collapse time, $t_{\rm c0}$,
given by the model in Ref. 2.
This figure shows that the collapse time
is accelerated, with respect to the model in ref. 2, because of
the energy acquired by the shells during encounters with unbound clumps.
In Fig.~4 we show  $t_{\rm c}$, for $\nu=2$. As in the previous figure the
solid line represents $t_{\rm c}$ taking account of the unbound clumps
effect, while the dashed line represents $t_{\rm c0}$ given by
the model in Ref. 2. Comparing Fig.~3 and Fig.~4 we see that the
effect of the unbound peaks on the acceleration of the
collapse time is larger in the case $\nu=1.2$. This is in agreement to
what observed for $\Delta E_{\rm T}$.\\
Finally, in Fig.~5 we make a comparison between the acceleration in
collapse time due to unbound clumps to the slowing down effect
produced by the gravitational
interaction of the quadrupole moment of the system with the tidal field of
the matter of the neighboring proto-clusters, described in
Ref. 3.

As shown in Fig.~5, the magnitude of
the two effects is comparable at $1 h^{-1} {\rm Mpc}$, but
the accelerating effect produced by the high speed clumps
decreases with increasing distance from the cluster centre,
%%%is a bit larger
%%In the region $(4  \div  6) h^{-1} $ Mpc
%%the effect produced by the high speed clumps is a bit larger
with respect to the slowing down effect produced by the tidal field. \\
%%This is probably
%%due to the rapid increase of high speed clumps with increasing distance. \\
The mechanism considered in this paper is different from that described in
Ref. 3 for two reasons:
\begin{itemize}
\item the delay of the collapse
produced by unbound clumps is due to encounters of infalling
shells of matter and unbound clumps. The effect described in Ref. 3 is
produced by tidal interaction of the matter of a given
protocluster with that of the neighboring ones;
\item the substructure responsible for collapse delay in this paper
is that contained inside the infalling shell while in Ref. 3 it is external
to the shell.
\end{itemize}
%%Altough
The two effects have a different physical origin,
%%%both mechanism
and act in the opposite direction: unbound clumps accelerate the
collapse, the effect of the tidal field is to delay
%%they delay
the shell collapse.
\begin{figure}
\psfig{file=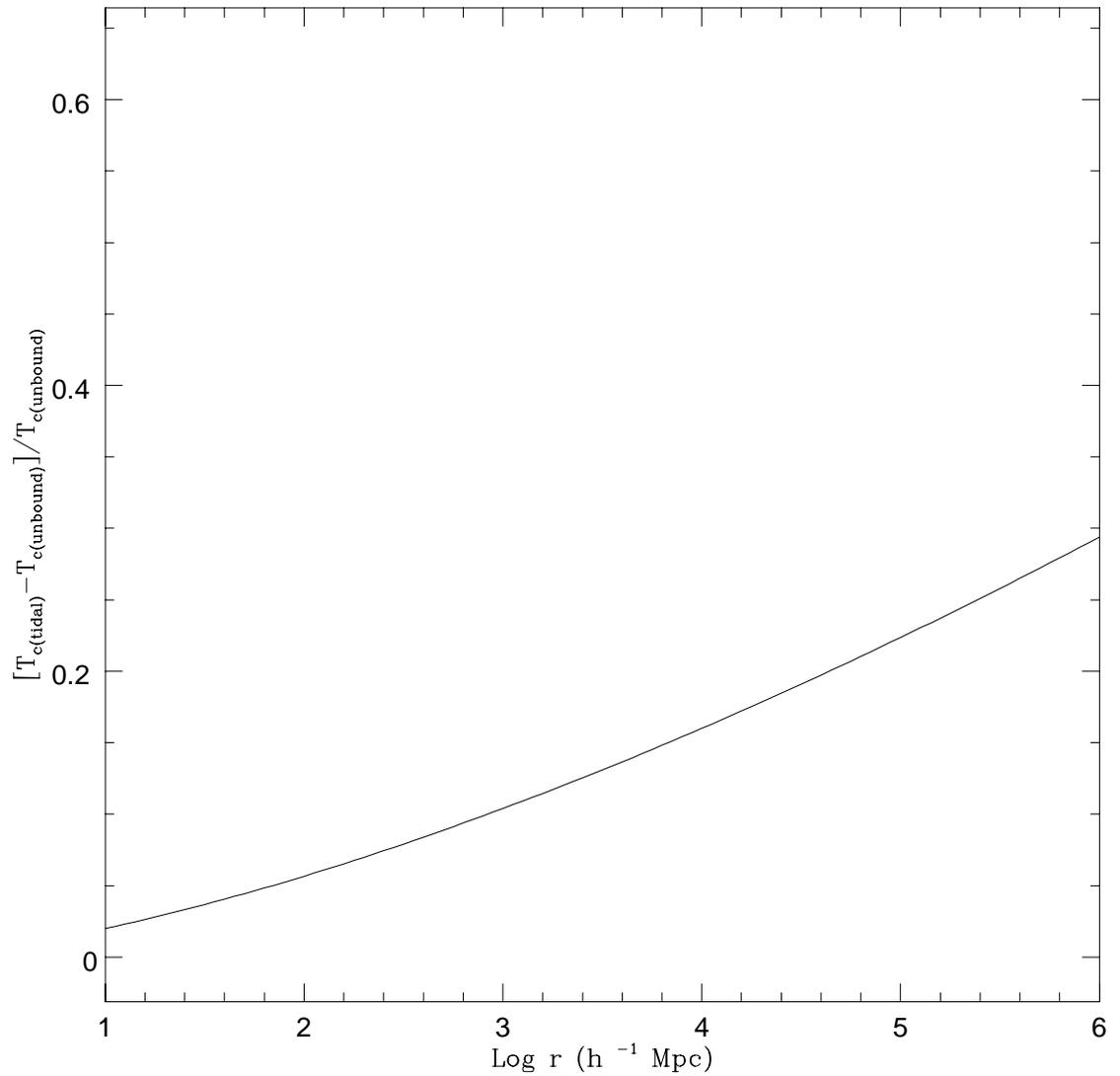,width=16cm}
\caption[]{Comparison of the collapse time $T_{\rm
c(unbound)}$, obtained taking account unbound high velocity
clumps, with $T_{\rm c(tidal)}$, obtained taking account of the
tidal interaction of the protocluster with the neighboring ones.
In this case $\nu=2$} \label{Fig. 5}
\end{figure}
It is obvious that during the infall of a shell both effects are present
with the final result that at small radii ($\simeq 1 h^{-1} {\rm Mpc}$) they
counterbalance while at larger radii tidal field is dominant producing a
slowing down of the shell collapse.
%%the collapse delay is even larger than that
%%described in this paper or in DG.
%%
%%%This means
%%%that a more realistic study of infall of matter and cluster formation
%%%should take account of both the two mechanisms.\\
%%
We want to stress that the idea that high velocity clumps can influence
the collapse and secondary infall of matter in protoclusters was
introduced (but not developed) in a paper by
Antonuccio \& Colafrancesco$^{37}$.
In the conclusion of that paper, regarding the velocity correlation tensor
and unbound substructure around density peaks, they noted that high
velocity clumps can exchange momentum with the collapsing or
infalling clumps and generate a drag force onto the collapsing material
with a consequent delay of the collapse. According to their paper
the drag force is due to dynamical friction. Unfortunately
this last considerations
%part of the idea
are not quite right because: \\
%At this point
%it is important to stress that
1) the effects of the high velocity clumps
on the collapse shell of matter cannot be calculated using the dynamical
friction approximation since a
necessary condition to apply this last is
%%this last can be used only in the case in which
that the encounter of two
systems is characterized by $V \simeq v_{\rm tip}$
(where $v_{\rm tip}$ is the internal velocity of the system)
(see Ref. 44).
To be more precise, according to
Ref. 49, a body of mass $M$ moving through a homogeneous
and infinite distribution of bodies of mass $m$, that we call field particles,
is subject
to the deceleration:
\begin{equation}
\frac{d {\bf v}_{\rm M}}{d t}= -16 \pi^2 \ln \Lambda G^2 m (M+m)
\frac{\int_0^{v_{\rm M}} f(v_{\rm m}) v^2_{\rm m} d v_{\rm m}}{v^3_{\rm M}}
{\bf v}_{\rm M}
\label{eq:dyn}
\end{equation}
where $\ln \Lambda$ is the Coulomb logarithm and $f(v_{\rm m})$ is
the phase-space number density field particles. As clearly stated
by equation~(\ref{eq:dyn}) only field particles moving slower than
$M$ contribute to the force. In our case, the role of the field
particles is played by the high speed clumps that have speeds much
larger than those characterizing the motion of the shell particles
(when IA is satisfied)
or at least a bit larger. \\
2) The effects of the high velocity clumps is to accelerate the shell
collapse and not to delay it. \\
%
%it is well
%known that dynamical friction is
%due only to the interaction of a test particle with field particles
%having velocity smaller than that of the test particle. Since the velocity
%of unbound clumps (field particles) are higher than that of the clumps
%in the shell (test particles), dynamical friction is ineffective..... \\
%
We note that the erroneous conclusion of Antonuccio \&
Colafrancesco$^{37}$ paper is not connected with their analysis of
the conditional probability distribution $ f_{\rm pk}({\bf v}|\nu)
$ developed in the paper, that we used to calculate the number of
subpeaks satisfying the IA. The erroneous conclusion is due to the
way they thought to calculate the interaction between the high
velocity clumps and the infalling shells.

\section{Conclusions}

In this paper we have studied the role of unbound high velocity
clumps on the collapse of
density peaks in a SCDM model ($\Omega_0=1$, $h=0.5$, $n=1$).
%solving numerically the equations of motion of a shell of
%barionic matter falling into the central regions of a cluster of galaxies.
We have shown that: \\
1) the encounters of collapsing shells with
unbound clumps produce an acceleration in the collapse of
density peaks. The effect is larger for peaks
having a lower value of $ \nu$ (see Fig.~2 $ \div $ 3). \\
2) A comparison between the collapse acceleration due to
encounters of shells with unbound clumps and the slowing down
effect produced by tidal interaction between a protocluster and
the neighboring ones (see Ref. 3) has shown
that these two effects act in opposite direction, namely the first
accelerate the shell collapse, while the second delay the
protocluster collapse, especially in its outskirts. The magnitude
of the two effects is comparable at $1 h^{-1} {\rm Mpc}$ but the
effect of the tidal field dominates at larger
radii. \\
3) The acceleration in the collapse is due to the additive action
of several high speed clumps with the matter of the infalling shells.
% Tolto
%%: the effect of
%%a single clump is negligible.
%
%%This corroborates the results obtained in some papers of ours
%%(DG; Del Popolo \& Gambera 1999;
%%Del Popolo et al. 1999a,b).

\section*{Acknowledgments}
%%%%%%Work partially supported by funds ex-60\% 98.
%%V.A.-D. would like to thank Prof. Giuseppe Moncada.
We are grateful to E. Spedicato, E. Recami, G. Mamon and O. Gnedin
for stimulating discussions during the period in which this work
was performed and whose comments and suggestions helped
us to improve the quality of this work.
A. Del Popolo and E.Nihal Ercan would like to thank
Bo$\breve{g}azi$\c{c}i University
Research Foundation for the financial support through the project code
01B304.\\

\end{document}